# 3D lattice Monte Carlo modeling of morphology formation of Si/SiOx nanocomposites during phase separation of nonstoichiometric Si oxide films

Ivan Oliinyk[1] and Andrey Sarikov[1-3]

[1] Educational Scientific Institute of High Technologies, Taras Shevchenko National University of Kyiv, 4-g Hlushkova Avenue, 03022 Kyiv, Ukraine

[2] V. Lashkaryov Institute of Semiconductor Physics, National Academy of Sciences of Ukraine, 41 Nauky Avenue, 03028 Kyiv, Ukraine

[3] National Technical University of Ukraine "Igor Sikorsky Kyiv Polytechnic Institute", 37 Beresteiskyi Avenue, 03056 Kyiv, Ukraine

**Abstract**

     In this paper, a three-dimensional lattice model based on the Monte Carlo approach is presented. This model is developed to investigate the kinetics of morphology change during phase separation in nonstoichiometric Si oxide ($SiO_x$, $x < 2$) films. The model takes into account the $SiO_x$ local atomic structure and probabilistic migration of oxygen atoms driven by the tendency of free energy minimization. The influence of the initial $SiO_x$ stoichiometry index $x$ and film thickness on the morphology of the precipitated Si phase in the Si oxide matrix is analyzed. The morphology of the Si phase is shown to critically depend on the initial $SiO_x$ stoichiometry. Namely, isolated Si nanoparticles form at low excess Si content ($x \geq 1.4$), while interconnected Si networks always appear at $x \leq 0.8$. A dimensional effect on the morphology of the Si phase is revealed. Namely, reducing the film thickness imposes geometric constraints on the Si network formation. The percolation threshold is found to shift from $x_p \approx 1.35$ for the bulk-like $SiO_x$ layers to $x_p \approx 0.85$ for the quasi-two-dimensional films. The transition to the bulk material behavior is observed at a $SiO_x$ thickness of approximately 4.2 nm.

## 1. Introduction

     Nanocomposite structures, in which Si particles are embedded in a dielectric matrix (primarily $SiO_2$), demonstrate unique luminescence and charge transport properties that make them promising particularly for light-emitting devices, high-efficiency solar cells, and non-volatile memory applications [1, 2]. The mostly used fabrication method of such nanocomposites is phase separation of nonstoichiometric Si oxide films ($SiO_x$, $x < 2$), which takes place during annealing at high temperatures (of the order of 1000°C) [3]. The performance of the Si/Si oxide nanocomposite based devices largely

rely on the morphology of the Si phase such as nanoparticle size, density, shape and spatial distribution within the Si oxide matrix. A particular importance for charge transport and charge storage applications is given to the percolation threshold of Si nanoparticles, i.e. the value of the stoichiometry index of the initial SiO$_x$ film, which delineates formation of connected Si networks and isolated Si particles [4, 5].

In modern technology, a transition from bulk films to spatially confined structures, such as ultrathin layers in nanodevices, is often required [6]. Experimental data evidence that reducing the SiO$_x$ layer thickness to the nanometer scale can significantly alter the phase separation process and the final morphology of the nanocomposite compared to bulk films [7]. Given the complexity of experimental research of the formation of ultrathin Si/Si oxide composites [8], there is a strong need for adequate modeling tools capable of predicting the kinetics and morphology of nanocomposites under geometric constraints.

Existing theoretical models to describe phase separation of nonstoichiometric Si oxides and related structures are often based either on thermodynamic calculations [9, 10], which cannot describe the Si nanoparticle shape, or on simplified kinetic schemes that ignore real atomic transport mechanisms [4, 11, 12]. It was demonstrated in [13] that the diffusion coefficient of Si is insufficient to explain the fast observed kinetics of the discussed process. Instead, an alternative approach based on oxygen migration provides the results more consistent with experimental observations. While two-dimensional Monte Carlo models based on the oxygen migration mechanism were implemented to estimate the percolation threshold in the Si/Si oxide systems [14], they fail to capture 3D effects including those related to the geometrical constraints of the SiO$_x$ films.

This work aims at developing a three-dimensional lattice model based on the Monte Carlo method to describe the phase separation kinetics in SiO$_x$ films and to investigate the influence of initial stoichiometry and film thickness on the Si phase morphology.

## 2. Model

The kinetics of phase separation in nonstoichiometric Si oxide was modeled on a three-dimensional lattice mimicking a diamond-like Si structure. Each node of this lattice contained a Si atom, and the midpoints between the nodes corresponded to possible oxygen atom positions. The lattice contained $N_x \times N_y \times N_z$ unit cells so that the total number of the Si nodes was

$$N_{Si} = 8\, N_x \times N_y \times N_z \qquad\qquad (1)$$

The thickness of the considered system $L_z$ was determined by the number of the layers along the $z$-axis $N_z$. An effective lattice constant $a_{eff} \approx 0.7$ nm was calculated based on the Si–O chemical bond length ($\sim 1.6$ Å) and the bond angle $\theta = 144°$ [15]. Hence, the thickness of the layer was $L_z \approx 0.7 N_z$ nm.

For simulations, an approach largely adopted from the two-dimensional model [14] was used. The initial SiO$_x$ film was formed by randomly placing $N_O = xN_{Si}$ atoms in the oxygen positions between the pairs of Si nodes in the lattice. The Si oxide microstructure obtained in this way could be considered as consisting of Si–O$_a$Si$_{4-a}$ ($0 \leq a \leq 4$) complexes with the central Si atom. Such description enables characterizing the free energy transformations during phase separation process in terms of the $a$-dependent complex penalty energy $\Delta_a$ ($\Delta_0 = \Delta_4 = 0$, $\Delta_1 = 0.5$ eV, $\Delta_2 = 0.51$ e, and $\Delta_3 = 0.22$ eV) [16]. During simulations with the kinetic Monte Carlo method, an oxygen atom capable of hopping to at least one vacant site around two neighboring Si atoms was randomly selected. The change in the free energy of the system as a result of the hopping event, $\Delta E$, was calculated as the difference in the total penalty energies after and before the hopping. The Metropolis algorithm [17] was applied to infer about success of the hopping event. Namely, this event was always successful and the oxygen atom moved to a new position if $\Delta E \leq 0$. At $\Delta E > 0$, a random number $0 \leq r \leq 1$ was generated and the hopping event was successful when the Metropolis criterion was satisfied:

$$r < \exp\left(-\frac{\Delta E}{k_B T}\right) \qquad (2)$$

Here, $k_B$ is the Boltzmann's constant and $T$ is the annealing temperature, respectively.

To quantify the formation of a continuous Si network (percolated Si phase) within the Si oxide matrix, the Largest Cluster Fraction (*LCF*) parameter was introduced as follows:

$$LCF = \frac{N_{max-cluster}}{N_{total-Si^0}} \qquad (3)$$

where $N_{max-cluster}$ is the number of Si atoms in the Si$^0$ oxidation state (having four neighboring Si atoms) belonging to the largest cluster found, and $N_{total-Si^0}$ is the total number of Si atoms in the Si$^0$ oxidation state in the entire system, respectively. The cluster size was determined using the Union-Find algorithm. The percolation threshold $x_p$ was defined as the initial SiO$_x$ stoichiometry index $x$ at which the *LCF* exhibited a sharp transition from approximately 0 (isolated particles) to approximately 1 (connected network).

## 3. Results and discussion

In this section, we present the results of the simulations on a three-dimensional lattice with the dimensions of $125 \times 125 \times N_z$, where $N_z = \{2, 5, 6, 7, 8, 12, 20\}$, over a stoichiometry range $x = 0.5\ldots1.7$ at the temperature of 1000°C. Each simulation was performed at the number of Monte Carlo steps up to $7 \times 10^9$. The simulated nanocomposite morphologies were visualized using an Open VIsualization TOol (OVITO) software [18].

Figure 1 shows the dependences of the *LCF* value for the phase-separated Si/Si oxide nanocomposite on the initial $SiO_x$ stoichiometry $x$ for each of the studied film thickness $N_z$. Analysis of these dependences allows us to draw the following conclusions. The *LCF* value is close to zero (~ 0.01…0.03) when the initial $SiO_x$ stoichiometry index $x \geq 1.4$. This indicates that at a low relative content of excess Si, nontoichiometric Si oxide decomposes forming isolated nanoparticles. At $x \leq 0.8$, $LCF \approx 1$, which points to formation of a continuous Si network. The transition between formation of isolated Si nanoparticles and percolated Si networks is dependent on the film thickness. As can be seen from Figure 1, at $N_z = 2$ ($L_z \approx 1.4$ nm), the percolation threshold is $x_p \approx 0.85$. This value has a good agreement with the value between about 0.7 and 0.8 obtained for a two-dimensional case [14]. Therefore, the behavior of the $SiO_x$ films with such thicknesses may be considered as a quasi-two-dimensional one. This conclusion is further suppoted by Figure 2, which shows that flat quasi-two-dimensional Si phase in the form of continuous Si network for $x = 0.8$ or isolated particles for $x = 0.9$ is formed for such $SiO_x$ film thickness value.

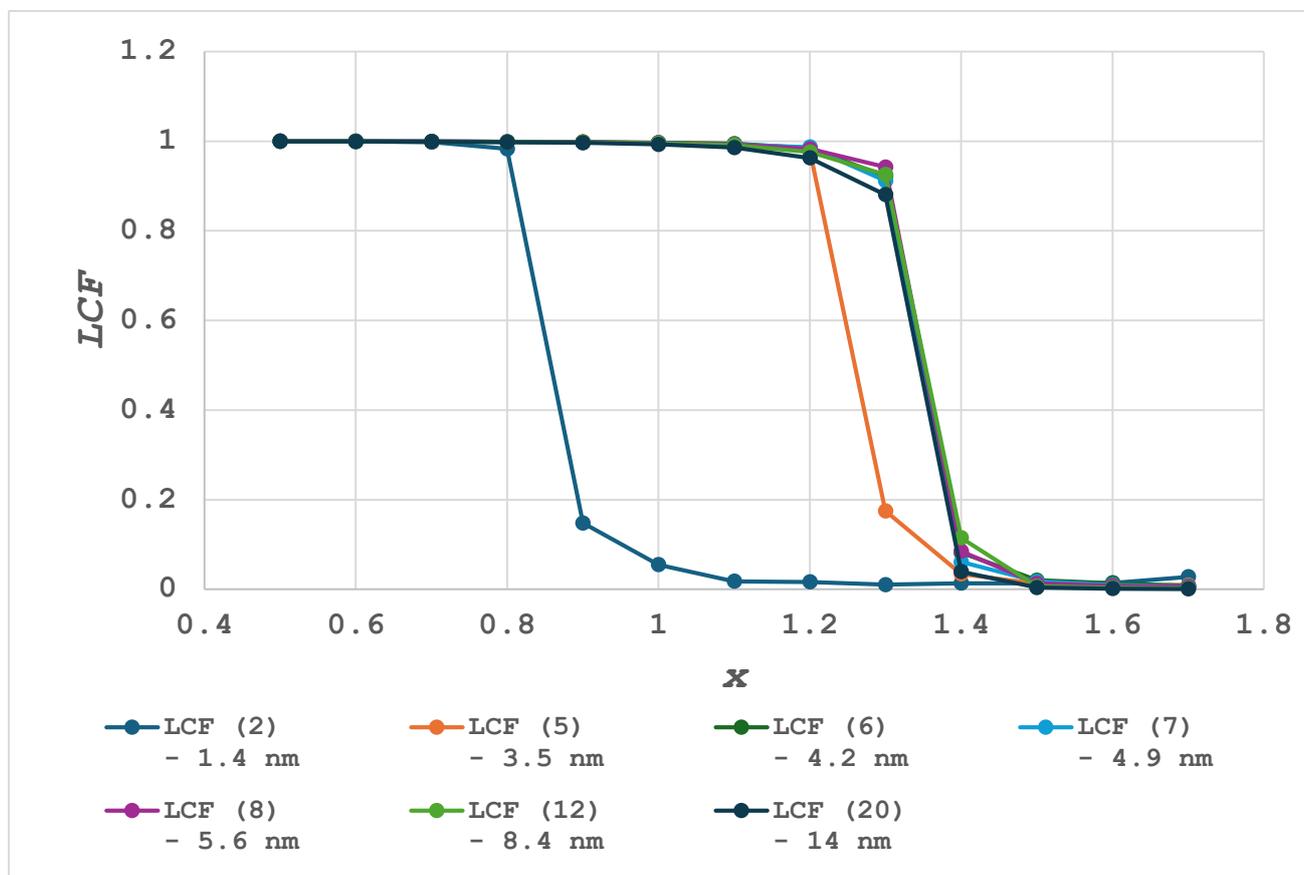

Figure 1. *LCF* versus initial stoichiometry of $SiO_x$ films with different values of $N_z$ corresponding to different film thicknesses.

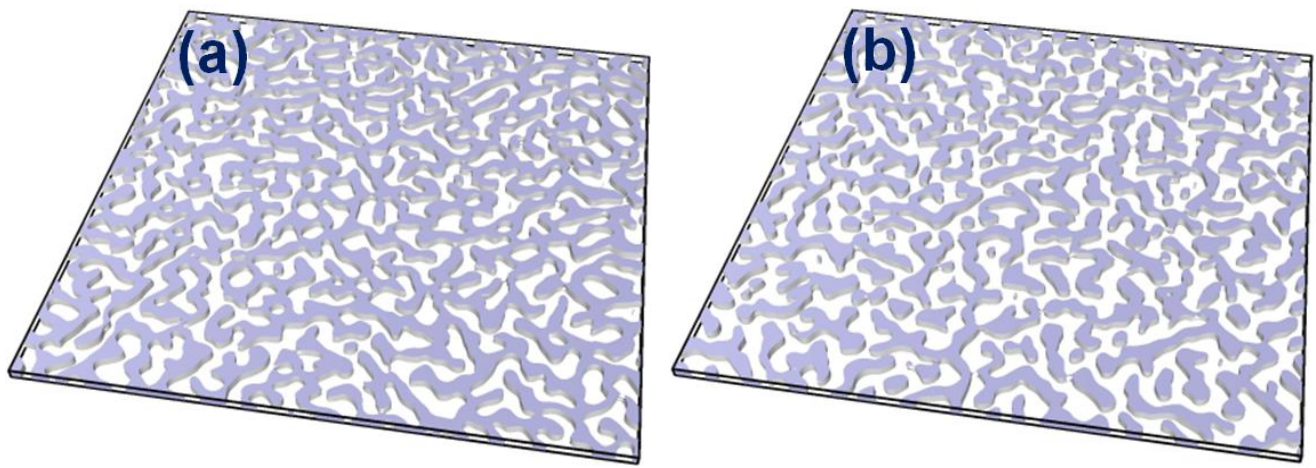

Figure 2. Simulated morphology of the Si phase in a quasi-two-dimensional system ($N_z = 2$): (a) $x = 0.8$, $LCF = 0.98$ – formation of a continuous but flat Si network, spreading mainly in the film plane; (b) $x = 0.9$, $LCF = 0.15$ – the system is just below the percolation threshold, and numerous isolated flat Si islands are formed.

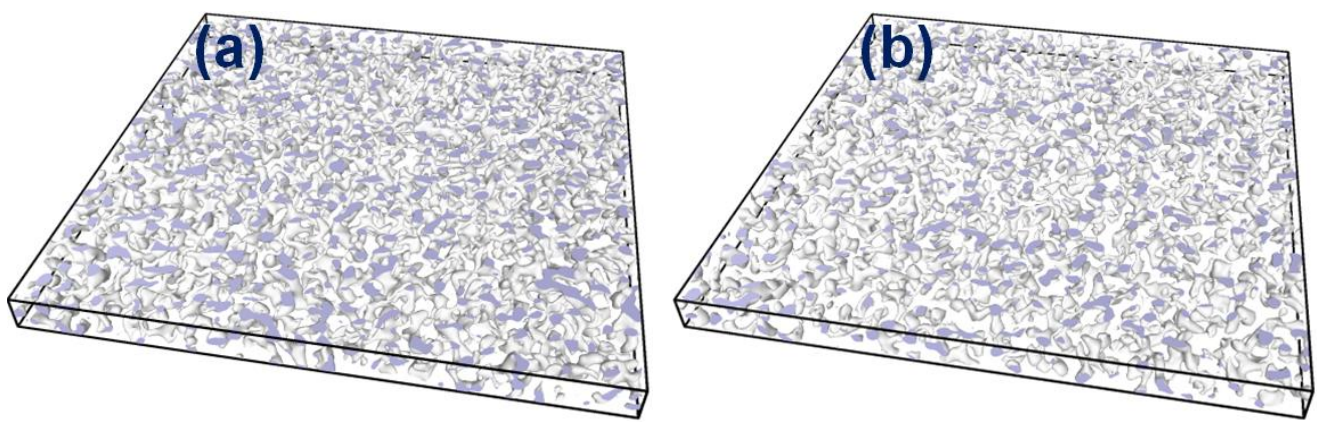

Figure 3 Simulated morphology of the Si phase for a three-dimensional system ($N_z = 8$): (a) $x = 1.3$, $LCF = 0.94$ – a complex, sponge-like three-dimensional Si network, which permeates the entire volume, is formed; (b) $x = 1.4$, $LCF = 0.08$ – the system is just below the percolation threshold, isolated Si nanoparticles, which have complex, non-spherical three-dimensional shapes, are formed.

Subsequent analysis of the data presented in Figure 1 demonstrates that raising the film thickness increases the value of $x_p$ due to appearance of off-plane connection paths for formation of already three-dimensional Si networks. This is supported by Figure 3, which shows formation of a three-dimensional Si network (Figure 3(a)) and three-dimensional isolated Si particles (Figure 3(b)) for the $SiO_x$ film thickness corresponding to $N_z = 8$ as an example. As can be further seen from Figure 1, the value of $x_p$ reaches saturation at approximately 1.35 at $N_z = 6$ (the film thickness of about 4.2 nm). This means that beginning from this thickness, the Si oxide film fully behaves as a three-dimensional object. As can be seen from Figure 1, further increase in the $SiO_x$ film thickness, e.g. to ~14 nm ($N_z = 20$) no longer changes the percolation threshold value.

Figure 4 presents visualization of the simulated structures with the initial stoichiometry value $x$ = 1.6 for $N_z$ = 2 and $N_z$ = 8 showing formation of isolated quasi-two-dimensional and purely three-dimensional Si nanoparticles. This supports the conclusion derived from the analysis of Figure 1 that no Si networks are possible to form at high excess Si contents independently on the SiO$_x$ film thickness.

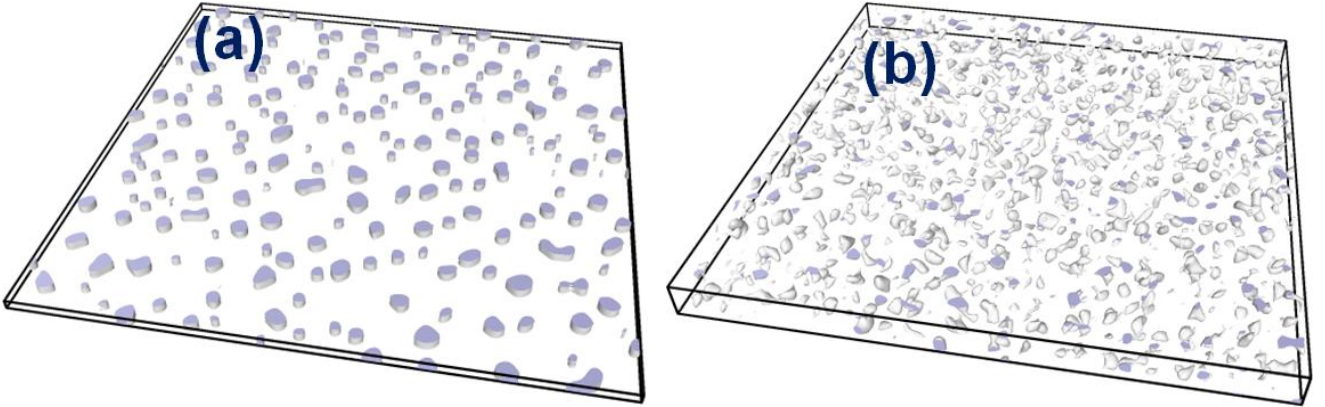

Figure 4. Simulated morphology of the Si phase at low excess Si content ($x$ = 1.6): (a) $N_z$ = 2 – quasi-two-dimensional case; (b) $N_z$ = 8 – three-dimensional structure.

## 4. Conclusion

In conclusion, a kinetic Monte Carlo model on a three-dimensional lattice was developed to simulate phase separation process in nonstoichiometric Si oxide films. The simulations allowed us to study the effect of the SiO$_x$ stoichiometry and film thickness on the formation of the morphology of the separated Si phase. The obtained results may be summarized as follows:

1. Depending on the initial SiO$_x$ stoichiometry, two types of the Si phase morphology may form, namely isolated Si particles at $x \geq 1.4$ and an interconnected sponge-like network at $x \leq 0.8$.

2. The percolation threshold of the Si phase critically depends on the SiO$_x$ film thickness. The value $x_p \approx 0.85$ for the thinnest films ($L_z \approx 1.4$ nm) and increases with the film thickness due to formation of off-plane connection pathways between the Si nanoparticles. This corresponds to a gradual transition from a quasi-two-dimensional to three-dimensional film behavior. A critical thickness of approximately 4.2 nm was determined, above which the SiO$_x$ film exhibits bulk material properties.

The obtained results give a deeper insight in understanding of the processes of formation of the morphology of Si/Si oxide composites by phase separation of nonstoichiometric Si oxide films as well as provide a theoretical basis for optimizing fabrication of composites with controlled morphology for specific electronic and photonic applications.